\documentstyle[prl,aps,multicol]{revtex}
\def\dto{(ET)$_4$\-[Ni(dto)$_2$]}
\def\ET{ET}
\def\cm{cm$^{-1}$}

\begin{document}
\draft
\title{Correlation gap in the optical spectra of the two-dimensional organic
metal (BEDT-TTF)$_4$\-[Ni(dto)$_2$]}
\author{P. Haas$^1$, E. Griesshaber$^1$, B. Gorshunov$^1$\cite{permanent1}, 
D. Schweitzer$^1$, M. Dressel$^1$\cite{email}, 
T. Klausa$^2$, W. Strunz$^2$, and F.F.  Assaad$^3$}
\address{$^1$ Physikalisches Institut, Universit\"at Stuttgart,
Pfaffenwaldring 57, D-70550 Stuttgart, Germany\\
$^2$ Anorganisch-Chemisches Institut, Universit\"at Heidelberg, 
Im Neuheimer Feld 270, D-69120 Heidelberg, Germany\\
$^3$ Institut f{\"u}r Theoretische Physik III, Universit\"at Stuttgart,
Pfaffenwaldring 57, D-70550 Stuttgart, Germany}
\date{Received \today}
\maketitle

\begin{abstract}
Optical reflection measurements within the highly conducting 
($a,b$)-plane  of the organic metal  (BEDT-TTF)$_4$\-[Ni(dto)$_2$] reveal 
the gradual development of a sharp feature  at around 200~\cm\ as  
the temperature is reduced below 150~K. Below this frequency a 
narrow Drude-like response is observed which accounts for the 
metallic behavior. Since de Haas-von Alphen oscillations at low 
temperatures confirm band structure calculations of bands crossing 
the Fermi energy, we assign the observed behavior to a 
two-dimensional metallic state in the proximity of a correlation 
induced metal-insulator transition. 
\end{abstract}


\begin{multicols}{2}
\columnseprule 0pt
\narrowtext

Low-dimensional metals have attracted considerable interest recently 
due to observed deviations from the simple Fermi-liquid behavior. By 
now the electronic properties of quasi one-dimensional Bechgaard 
salts are fairly well understood in terms of a Tomonaga-Luttinger 
liquid \cite{Dardel93,Dressel96,Moser98,Vescoli98}; but there exists 
only little agreement in the understanding of the quasi 
two-dimensional systems like organic or high-temperature 
superconductors \cite{remark3}. 

The two-dimensional organic BEDT-TTF
salts, where BEDT-TTF, abbreviated by ET, stands for 
bis\-ethylene\-di\-thio\--tetra\-thia\-ful\-val\-ene, are widely 
studied for their superconducting properties (which are still 
heavily debated \cite{Ishiguro90}); only a few investigations deal 
with the metallic state, which seems to be highly interesting 
because its vicinity to unconventional states like a Mott-Hubbard 
insulator or antiferromagnetism. Most of the systems consist of 
monovalent anions with each \ET\ molecule donating half an 
electron. The new charge transfer complex based on the electron donor 
\ET\ and the acceptor nickel\-bis(dithiooxalate) 
[Ni(dto)$_2$]$^{2-}$  was  proven to be metallic down to low 
temperatures \cite{Schiller00}. The aim of our investigation was twofold:
first to understand why the material does not become 
superconducting although it is highly ordered; second 
to study the effect of electronic correlations in this prototypical
two-dimensional metal. Here we report on the optical 
properties of \dto\ which strongly 
indicates that a correlation gap opens at low temperatures.

Single crystals of \dto\  grown by electrochemical methods   were as 
large as $2\times 3 \times 0.5$~mm$^3$. The crystal structure 
consists of \ET\ stacks along the $a$-axis where the ethylene 
groups are highly ordered below 200~K as known from X-ray analysis. 
This resembles the $\beta$-modification of the \ET\ salts with 
one-dimensional stacks of \ET\ mo\-lecules, in contrast to the 
$\kappa$-phase where pairs of organic molecules are rotated by 
90$^{\circ}$ with respect to each other \cite{Ishiguro90}. Due to 
S-S contacts of the \ET\ molecules in neighboring stacks, \dto\ 
forms a conducting ($a,b$)-plane. These layers are separated along 
the $c$-direc\-tion by sheets of the [Ni(dto)$_2$]$^{2-}$. Although 
the absolute value of the  conductivity perpendicular to the planes 
is about one to two orders of magnitude smaller \cite{Schiller00}, 
\dto\ exhibits a metal-like temperature dependence in all three 
directions (lower panel of Fig.~\ref{fig:figure1}), i.e.\ the  conductivity 
increases like $T^{-2}$ below 100~K.

Our samples were also characterized by electron spin resonance (ESR) 
experiments in the  temperature range from 300~K down to 4~K  
(Fig.~\ref{fig:figure1}) which 
yield an absolute value of the spin susceptibility of $6.4\times 
10^{-4}$~emu/mole at room temperature. It decreases continuously by 
a factor of three when the temperature is lowered with the strongest 
reduction in the temperature range $150~{\rm K}>T>100$~K indicating 
that the density of states might be reduced \cite{remark2}. The ESR 
linewidth $\Delta H$ exhibits a maximum just below 100~K. 
This behavior is seen in all three 
directions. For $T<80$~K the drop of $\Delta H$ and the constant 
spin susceptibility are in qualitative agreement with Elliott's 
prediction for an isotropic metal. Internal fields lead to a 
pronounced anisotropy as reflected in the $g$-values (not shown); 
they are basically independent of temperature: $g_a=2.0037$, 
$g_b=2.0048$, and $g_c=2.0110$. 

The Fermi surface of \dto\ as obtained by band structure calculations
\cite{Schiller00} (inset of Fig.~\ref{fig:figure1}) is 
similar to that of $\alpha$-(ET)$_{2}$\-(K\-Hg\-(SCN)$_{4})$ 
and to the typical Fermi surface observed in the $\kappa$-phase of 
\ET\ salts. It consists of a two-dimensional hole pocket 
($\alpha$-orbit) which covers about 14\%\ of the first Brillouin 
zone; as well as one-dimensional open trajectories with a flat 
dispersion which are less than half filled. These calculations
are fully confirmed by Shubnikov-de Haas and de Haas-van 
Alphen (dHvA) experiments at low temperatures \cite{Schiller00}.

The polarized optical reflectivity $R(\omega)$ of \dto\ was measured 
at $5~{\rm K}<T<300$~K for the electric field parallel to the $a$ 
and to the $b$ axes. Using two Fourier-transform spectrometers and 
a quasi-optical submillimeter spectrometer 
we covered the spectral range from 20~\cm\ up to 8000~\cm. We 
combined all results on $R(\omega)$, and after using appropriate 
low-frequency (Hagen-Rubens) and high-frequency ($\omega^{-2}$  
decay) extra\-polations finally performed a Kramers-Kronig analysis to 
obtain the optical conductivity $\sigma(\omega)$  and dielectric 
constant spectra.

The frequency dependence of the reflectivity and the corresponding 
conductivity are displayed in Fig.~\ref{fig:figure2} for both 
polarizations of the highly conducting plane. For the $b$-direction 
the plasma edge is located at somewhat higher frequencies as compared to the 
$a$-direction. In 
both directions no simple Drude response is observed. 
As the temperature is reduced below 150~K, pronounced new features 
develop in the low-frequency domain: a dip in $R(\omega)$ with a 
corresponding mode in $\sigma(\omega)$ around 200~\cm, which we 
ascribe to excitations across a gap. 
As discussed below, we assign the origin of this gap to correlation effects.
Below this gap, a narrow 
Drude-like peak appears at low temperatures. The position of the 
maximum is the same for both axes and does not change with 
temperature. The intensity (oscillator strength) increases linearly 
by about a factor of 10 when going from $150~{\rm K}$ to 50~K and 
stays constant below.

We now discuss possible scenarios for the understanding of our
experimental results. The major challenge is to reconcile two 
aspects of the experimental data. On one hand, 
the observed quantum oscillations \cite{Schiller00} are in complete 
agreement with calculations of the electronic band structure which 
do not take into account correlation effects. On  
the other hand, the optical conductivity shows signs of a 
gap. This situation stands in contrast to other 
materials such as (ET)$_2$\-AuBr$_2$ which also exhibits 
a low temperature gap in the conductivity spectrum at 130~\cm, 
the origin being assigned to a spin-density-wave ground state 
\cite{Ugawa95}. This magnetic instability, originating from nesting 
properties of the one-dimensional parts of the Fermi surface, leads 
to a correlation-induced reconstruction of the Fermi surface topology 
which does not match band structure calculations. This has been 
confirmed by dHvA experiments \cite{Pratt92}. In the case of \dto\ 
no indication of magnetic ordering was found at low temperatures, 
hence the formation of a spin density wave can definitely be ruled 
out. In particular the $g$-value does not show any shift in this 
range of temperature which would indicate the development of an 
internal magnetic field \cite{remark1}.  
No X-ray study of \dto\ has been performed  at 
$T<200$~K which would give a clear answer to whether a charge density
wave ground state is present.
However, we do not observe an abrupt change in 
any physical quantity and therefore we do not expect an 
increase in size of the cell as reported in typical charge density 
wave systems \cite{Gruner88}.  Furthermore, such transitions should 
show up in the Fermi surface topology.

Given the above, the most natural way of understanding the 
experimental data is in terms of a metallic state in the proximity 
of a quantum phase transition which partially or totally gaps the 
Fermi surface. The feature in $\sigma(\omega)$ at 200~\cm\ is 
interpreted as a precursor effect of this transition. As one 
approaches the transition (by changing the relevant microscopic 
parameters) spectral weight will be transfered from the Drude 
response to frequencies above the charge gap. The question is then: 
what is the nature of this transition and which microscopic parameter 
controls it?  

To illustrate this idea, we consider one-dimensional organics such 
as (TMTSF)$_2$PF$_6$. Along the chains the optical conductivity has 
very similar features to those of \dto: a gap feature at 200~\cm\ 
with a small Drude peak containing 1 \%\ of the spectral weight. 
The Bechgaard salts consist of quarter-filled chains which are 
insulating at low temperature due to umklapp processes 
\cite{Dressel96,Giamarchi91} and/or the combination of  
dimerization and strong Coulomb interaction \cite{Pedron94}. The 
metallic behavior is attributed to interchain hopping which 
effectively warps the one-dimensional Fermi surface  thus leading  
to doping  away form the ideal quarter-filled case. In this picture, 
the conductivity spectrum corresponds to that of a metallic state in 
the proximity of a  metal-insulator transition, the control 
parameter being the (dimensionality driven) doping away from 
quarter-filling \cite{Vescoli98}. The Fermi-surface of the \dto\ 
material has one-dimensional features; however, interchain hopping 
is a sizable fraction of the intrachain hopping, which explains the 
nearly isotropic dc conductivity within the ($a,b$)-plane.  Although 
tempting, an interpretation of the data in terms of strongly coupled 
chains seems out of reach.

Since the conductivity of \dto\ 
is roughly isotropic in the ($a,b$)-plane it 
is appropriate to start with a two dimensional model. The unit-cell 
contains four \ET \ molecules each donating  half an electron to 
the  acceptor. Band structure calculations reveal two conduction 
bands. In comparison to the energy scale of the gap, the other two bands 
lie well below the Fermi energy and  we will omit them. 
In real space this leads to a model with two 
$s$-orbitals per unit-cell each orbital accommodating one hole,  
i.e.\ half-filling.  The reduction to a two-band model may be 
explicitly carried out by noting that there is one hopping matrix 
element between the \ET\ molecules which dominates, thus defining a 
dimer. Each unit cell contains a pair of dimers, and the two 
conduction bands are well reproduced by considering only the 
antibonding combination of orbitals on the dimer.  A similar 
mapping  was carried out  for the $\kappa$-phase  of the \ET\ salts 
\cite{Kino95}.

Correlation effects
are  taken into account with a Hubbard $U$-term
which sets an energy cost $U$ for doubly 
occupied $s$-orbitals. Since we have precisely one hole  per orbital 
large values of $U$ lead to the localization of holes and hence a
Mott insulating state \cite{remark4}. The critical value of $U_c$ at which the 
metal-insulator transition occurs depends sensitively upon details of 
the band structure. In particular, if nesting is present, as in the 
prototype  single band two-dimensional half-filled Hubbard model 
with nearest neighbor hopping, $U_c = 0$ and the gap in a mean-field 
approach scales as $e^{-t/U}$ where $t$ is the hopping matrix 
element.  On the other hand,  when frustration is present thus 
prohibiting magnetic ordering, one expects $U_c$ to be a sizable 
fraction of the bandwidth. 

The antiferromagnetic insulating phase of the $\kappa$-phase salts
is naturally described within the above approach if one chooses $U > 
U_c$. For the present material it is tempting to argue that $U < 
U_c$. This qualitatively explains the experimental data: the 
system remains metallic at arbitrary temperatures and no phase 
transition is present. Since the Mott transition is a mass 
divergent transition, one expects the Fermi surface to remain 
essentially unaltered as a function of $U < U_c$. As mentioned above 
this is consistent with the dHvA experiments \cite{Schiller00}.  An 
indication of the mass enhancement due to proximity of the Mott 
transition  stems from the fact that in $a$ and $b$ directions  
respectively only 2\%\ and 4\%\ of the total spectral weight of the 
optical conductivity is contained in the narrow Drude-like mode.  
Comparing  those values to the  {\it large} Fermi surface observed 
by quantum oscillations infers an enhanced mass. 

The major problem within the above approach is the small energy 
scale of the gap  which is more than an order of magnitude smaller 
than the bandwidth. Approximate nesting properties of the Fermi 
surface will certainly reduce the value of the gap. A calculation of 
the spin and charge suceptibilities  within the H\"ukel 
tight-binding model shows only slight enhancements at wavevectors 
corresponding to nesting of the one-dimensional portions of the 
Fermi surface. 

Finally some remarks on the superconducting state which was not 
detected in \dto\ down to 0.06~K; the effect of the dimensionality
is not clear yet,
but it should be noted that the system is the least two-dimensional \ET\ conductor.
As argued  above, in \dto\ the
Coulomb repulsion is not strong enough to drive the system to a Mott 
insulator.  Given the phase diagram  of the $\kappa$-phase as a function of 
pressure \cite{Ishiguro90}, we can only speculate that reducing the pressure by chemical 
substitution will drive the system to a superconductor and ultimately to a Mott
insulator. 
This would agree with superconductivity observed
at quantum phase transitions in numerous systems
\cite{remark3,McKenzie98}. We also predict a similar scenario in
$\alpha$-(ET)$_{2}$\-(K\-Hg\-(SCN)$_{4})$; optical investigations are
in progress.

Thus we can conclude that the energy gap which develops in the 
optical conductivity at around 200~\cm\ for temperatures $T\leq 
150$~K in both directions of the highly conducting ($a,b$)-plane of 
\dto\ is due to electronic correlations. To reconcile the facts that 
the Fermi surface shows no correlation-induced gap and that there is 
no phase transition as the temperature is lowered, we have argued 
that the system is in the proximity of two-dimensional Mott 
metal-insulator transition. The narrow Drude-like contribution to 
the optical conductivity which is responsible for the metallic 
conductivity and quantum oscillations at low temperatures contains 
only of few percent of the spectral weight. The  effective mass of 
these carriers has to be enhanced.

We thank A. Darjushkin for assistance and  G. Gr\"uner, H. J. Koo, and M.H. Whangbo
for instructive  discussions.

\begin{figure}
\caption{\label{fig:figure1}
Upper frame: the ESR linewidth 
$\Delta H$ of (BEDT-TTF)$_4$\-[Ni(dto)$_2$]
as a function of temperature for the magnetic field oriented in the 
three different directions as indicated. The solid triangles (correpsonding
to the right axis) show the temperature dependence of the 
spin susceptibility obtained from the integrated ESR intensity; the open triangles for 
$T<20$~K (for the high conductivity values) are evaluated by
skin-effect corrected Dyson-shaped ESR-signal. 
Lower frame: the temperature dependence of the electrical conductivity of 
(BEDT-TTF)$_4$\-[Ni(dto)$_2$] along
the three crystal axes. The inset shows a cut through the Fermi surface (after
\protect\cite{Schiller00}). The areas shaded light gray correspond to
the $\alpha$ orbit, the dark gray area is the $\beta$ orbit.}    
\end{figure}

\begin{figure}
\caption{\label{fig:figure2}
(a) Frequency dependent reflectivity and 
(b) optical conductivity along the $a$-direction of 
(BEDT-TTF)$_4$\-[Ni(dto)$_2$] for
different temperatures  as indicated.
(c) and (d) show the reflectivity and conductivity, respectively,
measured along the $b$-direction for different temperatures. In both orientations
a gap-feature is clearly seen at around 200~\cm.}
\end{figure}

\end{multicols}

\begin{thebibliography}{99}
\bibitem{remark4} The remnant $\ln(2)$ entropy per charge 
carrier associated with the spin degrees of freedom is in general 
lifted by magnetic ordering.
\end{thebibliography}
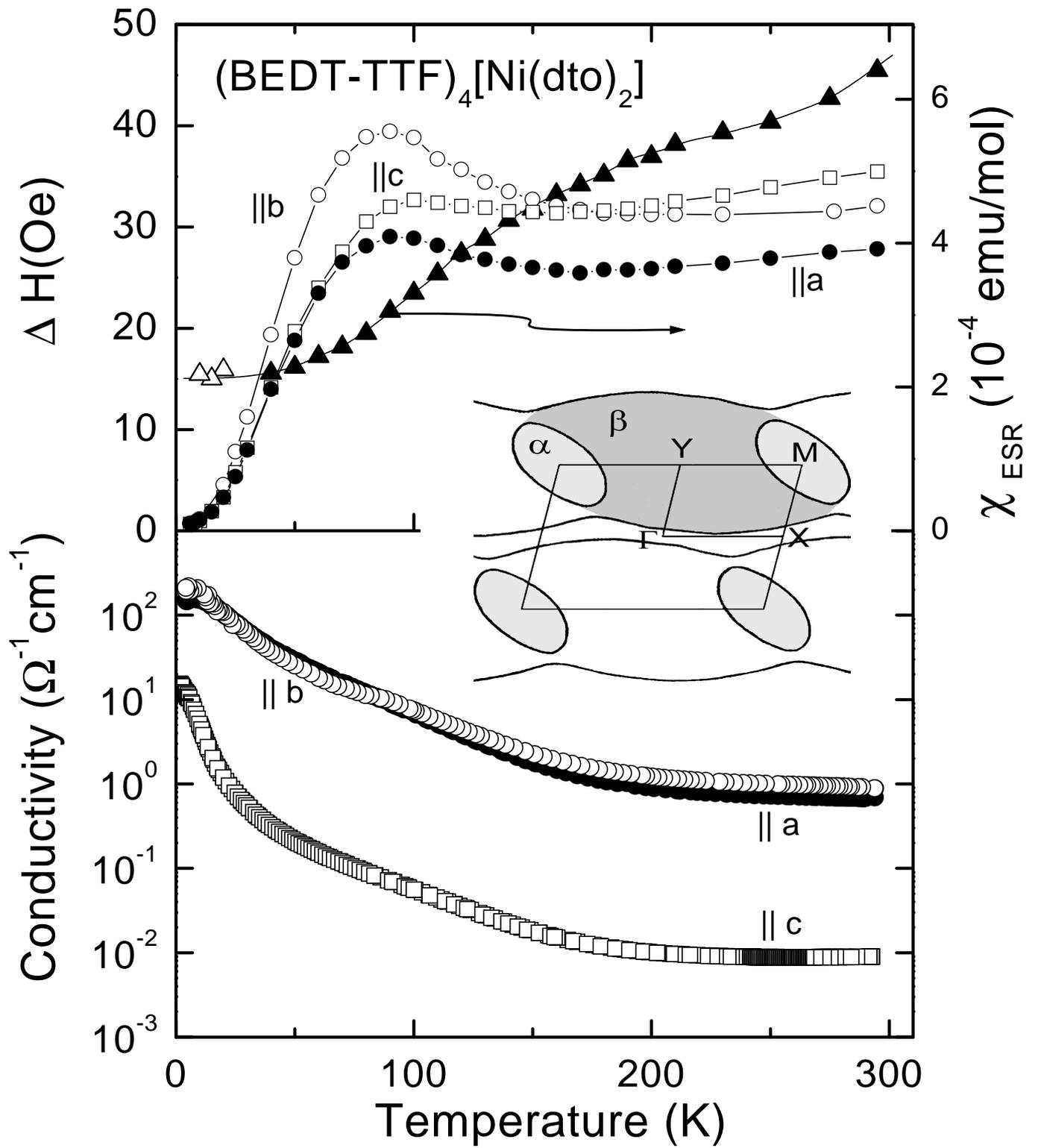
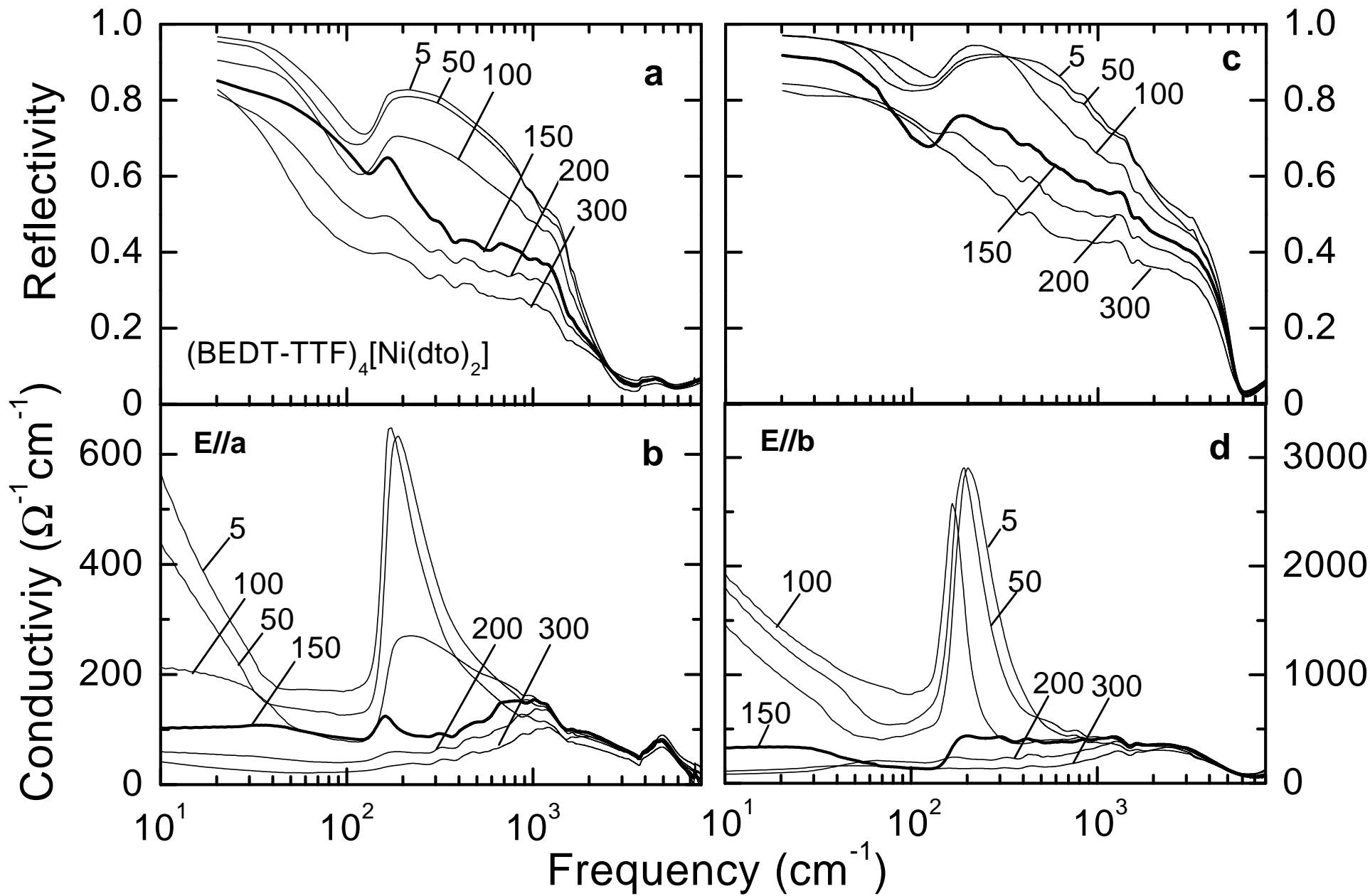
\end{document}